\documentclass[useAMS,usenatbib]{mn2e}

\usepackage[utf8]{inputenc}
\usepackage{amsmath}
\usepackage{amsfonts}
\usepackage{amssymb}
\usepackage{graphicx}
\usepackage{float}
\usepackage{natbib}
\usepackage{xcolor}
\usepackage{import}
\newcommand{\bL}{\mathbf{L}}
\newcommand{\ee}{\end{equation}}
\newcommand{\bea}{\begin{eqnarray}}                                              
\newcommand{\eea}{\end{eqnarray}}

\newcommand{\bl}{{\pmb \ell}}
\def\simlt{\lower.5ex\hbox{$\; \buildrel < \over \sim \;$}}
\def\simgt{\lower.5ex\hbox{$\; \buildrel > \over \sim \;$}}
\def\lya{Ly$\alpha$}
\def\lyb{Ly$\beta$}

\newcommand{\degree}{^{\circ}}

\def\aap{A\& A}

\def\apjl{ApJL}
\def\apjs{ApJS}

\title[Lyman-$\alpha$ Lensing]{Noise Estimates for Measurements of Weak Lensing from the Lyman-alpha Forest}

\author[Metcalf et al.]{R. Benton Metcalf$^{1,2}$\thanks{E-mail: robertbenton.metcalf@unibo.it}, Rupert A.C. Croft$^{3}$\thanks{E-mail: rcroft@cmu.edu} and Alessandro Romeo$^{1}$\\
$^1$ Dipartimento di Fisica e Astronomia, Universita di Bologna,
 viale B. Pichat 6/2, I-40127 Bologna, Italy\\
$^2$ INAF-Osservatorio Astronomico di Bologna, via Ranzani 1, 40127 Bologna, Italy \\
$^3$ McWilliams Center for Cosmology, Department of Physics, 
Carnegie Mellon University, Pittsburgh, PA 15213, USA\\
}
\begin{document}

\date{Accepted . Received ; in original form }

\pubyear{2017}

\maketitle

\label{firstpage}

\begin{abstract}
We have proposed a method for measuring weak lensing using the Lyman-$\alpha$ forest.  Here we estimate the noise expected in weak lensing maps and power spectra for different sets of observational parameters.  We find that surveys of the size and quality of the ones being done today and ones planned for the future will be able to measure the lensing power spectrum at a source redshift of $z\simeq 2.5$ with high precision and even be able to image the distribution of foreground matter with high fidelity on degree scales. For example, we predict that Lyman-$\alpha$ forest lensing measurement from the Dark Energy Spectroscopic Instrument survey should yield the mass fluctuation amplitude with statistical errors of 1.5\%. By dividing the redshift range into multiple bins some tomographic lensing information should be accessible as well.  This would allow for cosmological lensing measurements at higher redshift than are accessible with galaxy shear surveys and correspondingly better constraints on the evolution of dark energy at relatively early times.
\end{abstract}


\section{Introduction}

The study of the \lya\ forest  in the spectra of high redshift quasars and galaxies has been a great success for cosmology and galaxy formation.  It has been used to constrain the mass of neutrinos 
\citep{2015JCAP...11..011P,1999PhRvL..83.1092C}  and study the ionization of the intergalactic medium over comic history (see \citet{2016ARA&A..54..313M} for a review).  It has recently become possible to map the three dimensional distribution of HI at redshifts of $z\simeq 2.4$ using the \lya\ forest \citep{2014MNRAS.440.2599C,2001MNRAS.326..597P} and measure  Baryon Acoustic Oscillations (BAO) \citep{2013A&A...552A..96B,2013JCAP...04..026S}.

Weak Gravitational lensing has also provided important cosmological and astrophysical information in recent years and is expected to be an essential probe of dark energy and dark matter in the future \citep[for a review][]{2008ARNPS..58...99H}.  The power of weak lensing for measuring dark energy derives primarily from being able to measure the evolution of structure formation which is directly related to the evolution of the energy density in the Universe.  This is done by measuring the lensing power spectrum as a function of source redshift.  Secondarily, dark energy affects the angular size distance - redshift relation.  Extending the source redshift range beyond those of galaxies ($z\sim 0.4 - 1.4$) would allow for further exploration of dark energy and alternative gravity model space.  Weak lensing of the CMB is one way of doing this for a source redshift of $z \sim 1,000$ \citep{2016A&A...594A..15P} although it is relatively low signal-to-noise and has no range in source redshift.  It would be helpful to fill in this gap in source redshift with high signal-to-noise measurements.

The authors recently proposed that weak gravitational lensing could be measured with the \lya\ forest \citep{CRM17}.  A noisy mapping of the HI distribution at high redshift makes it possible to measure lensing by detecting the distortion of the angular correlation function, or power spectrum, from its expected isotropic form.  This is very similar to the methods used to measure lensing with the Cosmic Microwave Background (CMB) \citep{2006PhR...429....1L} and to proposed methods for measuring weak lensing with 21~cm radiation from high redshift \citep{metcalf&white2006,ZandZ2006,2015MNRAS.448.2368P}.  In this paper, we will make this proposal more concrete by estimating the noise levels with which we can expect to measure lensing in realistic \lya\ surveys.  In a more technical paper to follow we will show explicitly how a lensing estimator can be constructed in this case.

Future \lya\ lensing maps could be cross-correlated with galaxy shear and CMB lensing maps to get as large a range in source redshift as possible for probing the evolution of dark energy and modifications to General Relativity.  They could also be cross-correlated with foreground galaxy surveys to measure bias and directly measure the angular size distance redshift relation.

In the next section we review some of the proposed and ongoing \lya\ surveys and approximate some of the parameters of these surveys that will be relevant for lensing.  In Section~\ref{sec:formalism} we derive the  noise estimation in a lensing measurement from \lya\ data.  We show how to calculate the expected signal in Section~\ref{sec:signal} and then present estimates for particular sets of survey parameters in Section~\ref{sec:estimates}.  Section~\ref{sec:conclusion} contains a conclusion and discussion of future prospects.

\begin{table*}

\begin{tabular}{|llrrl|l|l|l}

\hline

Dataset   & When      & Area            & N$_{\rm spectra}$ & mean separation & mean S/N in flux\\ \hline

BOSS DR12 & 2016      & 10,000 sq. deg. & 160,000            & 15 arcmin       &  2.72\\

eBOSS     & 2014-2018 & 7,500 sq. deg.  & 270,000            & 10 arcmin       &  2.99 \\

CLAMATO   & 2014-2018 & 0.8 sq. deg.    & 1,000              & 1.7 arcmin      &  3.49 \\

WEAVE     & 2018-2025 & 6,000 sq. deg.  & 400,000            & 7.5 arcmin      & 3.26 \\

DESI      & 2018-2023 & 14,000 sq. deg. & 770,000            & 8.1 arcmin     &  3.26\\

Subaru PFS & 2019-2022  & 15 sq. deg. &  7,400   & 2.7 arcmin & 2.70\\ 

MSE       & 2025-     & 1,000 sq. deg.  & 1,000,000          & 1.9 arcmin  & 3   \\\hline

\end{tabular}
\caption{
Some relevant parameters for future \lya\ forest observational
datasets (see Section 2 for details). 
\label{table:surveytab}
}
\end{table*}

\section{Observational data samples}
\label{sec:surveys}

Lensing of the \lya\ forest can be carried out with samples of backlights
which are quasars, galaxies, or a combination of the two. In Table~\ref{table:surveytab} we have listed some currently available \lya\ forest
datasets, as well as some for which data collection is ongoing. We also list
some planned surveys and a possible survey with a proposed instrument
(Mauna Kea Spectroscopic Explorer, MSE\footnote{http://mse.cfht.hawaii.edu}). 
Surveys targeting the \lya\ forest from the ground are limited by 
atmospheric transmission to redshifts $z > 1.9$ and by the steep 
decline in the number of backlights to redshifts below $z = 3-4$. In this 
paper we will assume that the mean redshift of \lya\ forest pixels
is $z=2.5$. We will also make the assumption that all sightlines
in our sample fully cover the redshift interval $\Delta z$ we are
considering. Although the redshift range from the \lya\ line to \lyb\ line
is $\Delta z=0.55$ at $z=2.5$, the backlights will not all be at the same 
redshift, meaning that some spectra will only offer partial coverage.
These additional complications will be addressed in future work.

We estimate the mean signal-to-noise ratio for the various observational
samples (current and future) listed in table~\ref{table:surveytab}.
In order to
standardize our results, we assume that all observations are rebinned
into pixels with redshift extent $\Delta z=0.001$ (we ignore correlations
between pixels, so that the $S/N = \sqrt{0.001/(\Delta z)_1}$,
where $(\Delta z)_1$ is the original pixel size.
We also assume a mean redshift of
\lya\ forest observations of $z=2.5$. 

In the table we
give the S/N for the observed \lya\ forest transmitted flux in the spectra. 
If $F$ is this transmitted flux, the quantity
$\delta_{F}=(F/\langle F \rangle) -1$, can be related to the
underlying density field using a linear theory biasing relation.
The 3D power spectrum of \lya\ forest $\delta_{F}$ fluctuations
 is given by (McDonald 2003):
\begin{equation}
P_{\delta F}({\bf k}) = b^{2} (1+\beta\mu^2_k)^2 P_\delta(k),
\label{mcdonaldpk}
\end{equation} 
where $P_\delta(k)$ is the underlying dark matter power spectrum
and $\mu_k$ is the cosine of the angle between the
wave vector ${\rm \bf k}$ and the line of sight $\hat{z}$.  
$b$ and $\beta$ are parameters that describe the relative bias
between flux and matter fluctuations and the
strength of redshift distortions respectively.
In the case of galaxies, which are conserved under
redshift distortions, 
$\beta = \frac{d \ln D}{d\ln a} \simeq \Omega_m(z)^{0.55}$  where $D$ is the
linear growth rate \citep{1987MNRAS.227....1K}. For the \lya\ forest, which 
undergoes a non-linear transformation $F=e^{-\tau}$ between 
flux $F$ and optical depth $\tau$ (which is conserved),
$\beta$ is a separate parameter. At the redshifts $z=2-3$ of interest,
$b$ is approximately $0.2$ and $\beta$  unity \citep{2011JCAP...09..001S}.
As a result, the S/N in the density field will be approximately
equal to $b(1+\beta)\sim 0.3$ times the S/N in the observed
\lya\ forest flux \citep{2011JCAP...09..001S}, although it is the S/N
in the observed flux, given in Table~\ref{table:surveytab},  that is relevant for our estimates of lensing reconstruction.
Most of the flux S/N values in Table~\ref{table:surveytab} are $\sim 3$.

Simulations of the intergalactic medium predict that the transverse HI power spectrum will be damped on small scales by pressure \citep{2010MNRAS.404.1295P,2010MNRAS.404.1281P}.   At the redshifts of relevance the Jean's scale is $R_J \simeq 0.5$~Mpc.  In modeling the flux power spectrum in what follows we include a damping factor of $e^{-(k/R_J)^2}$ although this does not have a strong effect on the lensing results.

\citet{2013AJ....145...69L} give details for the BOSS DR9 \lya\
forest sample. We make the assumption that the signal-to-noise ratio
of this sample is similar to that in the final BOSS Data Release.
The spectra have been rebinned into vacuum wavelength pixels
of size $\Delta \log_{10}(\lambda)=10^{-4}$, i.e. $\Delta v=69.02$ km\,s$^{-1}$.
The median S/N per pixel values for the quasars in the sample are shown in
Figure 3 of Lee et al. Integrating we find that mean S/N for the flux
averaged over all quasars in $\Delta v=69.02$ kms$^{-1}$ pixels is 2.40.
In $\Delta z=0.001$ pixels this is equal to S/N=2.72.

The successor to BOSS, eBOSS \citep{2016AJ....151...44D} is ongoing, and 
the quasar survey will cover a smaller sky area, planned to 
eventually be three-quarters of the BOSS area.  The majority of
eBOSS quasars are at lower redshifts than those for which the
\lya\ forest is accessible, but there is also a large subset of quasars
with $z>2.1$. These \lya\ forest eBOSS targets are in two
categories: (i) quasars selected using an improved algorithm from those in 
BOSS, which uses more information on  quasar variability, and (ii) 
re-observation of the faintest BOSS quasars to increase their signal-to-noise.
The quasars in group (i) have to a good approximation the same S/N
as the BOSS quasars. For quasars in group (ii), we assume that the S/N ratio of
the lowest 50\% of BOSS quasars will be increased by a factor of approximately
$\sqrt{2}$. Coaddition of eBOSS and BOSS spectra is needed to achieve this.
Taking both of these factors into account the mean S/N in $\Delta z=0.001$
is 2.99.

Other surveys covering large fractions of the sky and which will be
starting soon are DESI \citep{2016arXiv161100036D} and WEAVE 
\citep{2012SPIE.8446E..0PD}. Both of these involve fiber 
spectrographs which are more highly multiplexed than BOSS, and this will
allow the number of \lya\ forest quasars observed to reach more than a 
million. With the current observing parameters, the S/N per unit 
wavelength in WEAVE and DESI is expected to be 50\% higher than
BOSS for objects with the same magnitude (M. Pieri, private communication)
WEAVE will include  a deep QSO survey (to $g<23.2$) covering 6000 sq. deg.
with a minimum S/N per \AA\ of 0.4, but likely covering a
limited redshift range ($\delta z \sim 0.5-1.0$). A bright QSO survey 
($g <20$) covering all $z>2.15$ quasars over 10000 sq. deg. will also 
be part of WEAVE, and these brightest will have mean S/N  per \AA\ of $\sim 7$. Other 
components of WEAVE include deep exposures of QSOs in the HETDEX fields
(450 deg. sq., leading to S/N per \AA\ of $> 4$). As WEAVE has many science
goals, including BAO measurement, this leads to many subsamples, some of
which are not yet defined fully. Accounting
for the full complexity of these surveys is beyond the scope of the present
paper, as we limit ourselves to a single S/N estimate and number of QSOs. 
Guided by the BOSS results we assume mean S/N per  $\Delta z=0.001$ of 
$2.72\times 1.5$ for the first 160,000 objects in WEAVE and mean S/N of $2.72$
for the rest, leading to an averaged S/N of 3.26.
DESI will cover a larger area than WEAVE, again going deeper than BOSS (
it will cover $r<23$), and with S/N threshold of 0.5.  We make the 
simplifying assumption that the S/N for DESI will be similar to WEAVE.

The surveys mentioned so far are all exclusively using QSOs to 
backlight the \lya\ forest. Because of their scarcity this limits the
sightline density to be relatively low, with mean separation of the
order of 10 arcmins.  \citet{2014ApJ...795L..12L} showed however that it is possible
to map out the \lya\ forest well using star forming galaxies as backlights.
Assuming that large aperture telescopes are available to observe these
fainter objects this greatly increases the potential sightline density. 
The CLAMATO survey\footnote{http://clamato.lbl.gov}
 is currently ongoing, aiming to cover a $\sim 1$~deg.~sq. portion of the  
COSMOS \citep{2007ApJS..172....1S} field, with
$\sim 1000$ \lya\ forest sightlines, mostly galaxies. From observations
completed so far, the  S/N level in the data obeys
$\frac{dn}{d(S/N)} \propto (S/N)^{-2.9}$ (K.G. Lee, private
communication). A minimum S/N cut of $S/N > 1.5$ per \AA\ is applied,
which leads to a mean $S/N$ of 3.49 in $\Delta z=0.001$ intervals.

The Prime Focus Spectrograph on the Subaru Telescope will be a powerful
resource for high redshift spectroscopy \citep{2012jgrg.conf.1502T}. High redshift galaxies will 
be suitable targets for a PFS \lya\ forest survey, and in this case
the S/N level in the data is likely to obey 
$\frac{dn}{d(S/N)} \propto (S/N)^{-3.6}$ (K.G. Lee, private
communication). As a result with a $S/N > 1.5$ cut, the mean $S/N$
is $\Delta z=0.001$ intervals is 2.7. The wide field of view of PFS will enable
sky coverage of 15 sq. deg. for a possible survey.

Surveys such as CLAMATO have shown that, 
the space density of galaxies at  redshifts $z=2-4$ is high enough
to map the \lya\ forest with close to arcminute angular resolution. 
Given enough resources, future surveys could exploit this over large
sky areas. Our final entry in Table~\ref{table:surveytab} is a more speculative
one along these lines. The Mauna Kea Spectroscopic Explorer (MSE) is a proposed
10m class multiobject spectroscopic facility. With dedicated use, such a 
telescope could be expected to make a million sightline survey possible
at such arcminute resolution. We assume a mean S/N ratio for the \lya\ forest 
flux of 3 to be
comparable to CLAMATO and PFS.

\section{estimating the noise in the lensing}
\label{sec:formalism}

Quasars and galaxies will be randomly, not regularly, distributed across the sky and distributed in redshift.  We will present a lensing estimator that deals with these complications in a more technical paper \citep{MCR3}, but here we will estimate the noise by imagining the sources are distributed in a regular rectangular grid on the sky and all at a redshift beyond the range considered.  In this case we can borrow some of the techniques used for lensing of the CMB and 21~cm emission without significant modification \citep[see][]{2001ApJ...557L..79H,ZandZ2006}.

The noise in the gravitational potential is given by
\begin{eqnarray}
\label{eq:Nellj}
N_\phi(L) = 
\left[\sum_j^{j_{\rm max}} \int_{\ell_{\rm min}}^{\ell_{\rm max}}
  \frac{d^2\ell}{(2\pi)^2} \frac{[\bl \cdot \bL C_{\ell,j}+\bL \cdot
    (\bL-\bl) C_{|\ell-L|,j}]^2}{2 C^{\rm tot}_{\ell,j}C^{\rm
      tot}_{|\bl-\bL |,j}}\right]^{-1}, 
\end{eqnarray}
where
\begin{equation}
\label{eq:Cellj}
C_{\ell,j} = \frac{P_{\delta F}(\sqrt{(\ell/{\cal D})^2+
(j2\pi/{\cal L})^2})}{ {\cal D}^2 {\cal L}}
\end{equation}
is an approximation of the the angular/radial power spectrum of the flux.
The total power spectrum in the denominator of (\ref{eq:Nellj}) is given by
\begin{equation}
C^{\rm tot}_{\ell,j} = C_{\ell,j} + C^N_{\ell,j}
\end{equation}
where $C^N_{\ell,j}$ is the power spectrum of noise from errors in the density estimate in each pixel.  The discrete form of this estimator and noise (on a rectangular grid) is also derived in \citet{2015MNRAS.448.2368P}.

The smallest angular scale that can be probed is set by the density of back lights on the sky.  
$\ell_{\rm max} \simeq \pi/\delta\theta$  where $\delta\theta$ is the average angular distance between back lights.  The largest scale is set by the dimensions of the survey $\ell_{\rm min} \simeq A_{\rm survey}^{-1/2}$ where $A_{\rm survey}$ is the area of the survey in steradians.  The maximum number of radial modes, $j_{\rm max}$,  is set by the number of pixels used in each of the spectra.

If $\sigma_\tau$ is the fractional error in the estimate of the absorption within one pixel then the noise in the angular density power spectrum is
\begin{align}
C^N_{\ell,j} & \simeq  \frac{1}{{\cal D}^2 {\cal L}} \frac{V}{N} ~\sigma_\tau^2 \simeq \frac{A_{\rm survey}\sigma_\tau^2}{j_{\rm max} (l_{\rm max} - l_{\rm min})^2 }  \\
& \simeq \frac{\sigma_\tau^2}{j_{\rm max}\, (l_{\rm max} - l_{\rm min})^2 \, l_{\rm min}^2}
\end{align}
where $N$ is the total number of data points and $V$ is the total volume within which the HI absorption is being measured.  If there are correlations between pixels then $C^N_{\ell,j} $ would have a $j$ dependence.  
These could easily arise from the fitting of the continuum.  We will ignore this complication here.

An estimate for the error in the band power spectrum is derived in the appendix and can be expressed as
\begin{align}
\Delta C_D(L) = \sqrt{\frac{4\pi}{A_{\rm survey} L \Delta L} } ~\left[ C_D(L) + N_D(L)  \right]
\label{eq:DeltaC}
\end{align}
 where $\Delta L$ is the width of the band.  The first term of this is the sample variance and the second comes from noise in the lensing measurement.
The estimated signal-to-noise for the power spectrum normalization, denoted $\mathcal{S}$, is also derived in the appendix.  This will be a convenient way to summarize the total signal-to-noise for a survey and for determining how well the linear growth factor can be measured.

\section{the lensing signal}
\label{sec:signal}

\begin{figure}
 \includegraphics[width=\columnwidth]{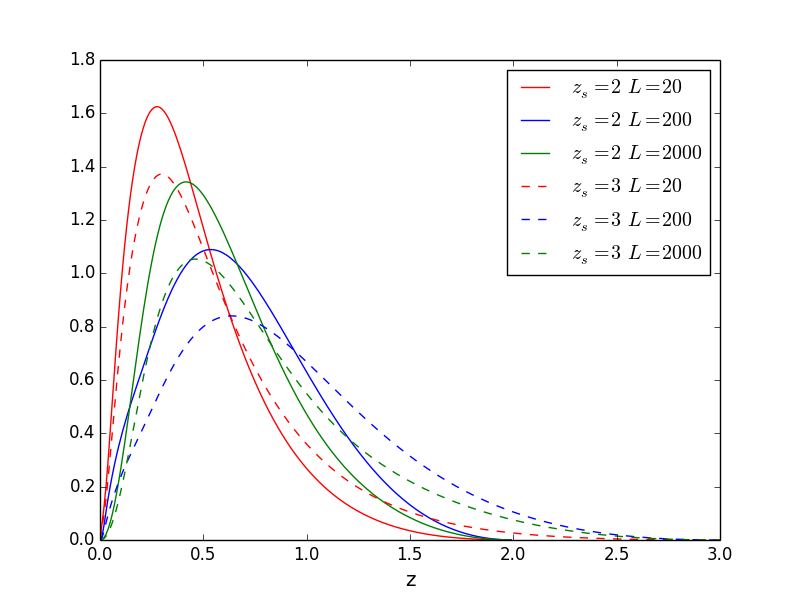}
 \caption{The contribution to the lensing power spectrum as a function of redshift.  Each curve is normalized to one.  The legend shows the source redshift and the multipole number.}
 \label{fig:dC_dz}
\end{figure}

The lensing potential is related to the density contrast along the line of sight in the weak lensing  Born approximation by 
\begin{align}
\kappa(\btheta, z_s) = \frac{3}{2} \frac{H_o^2 \Omega_m}{c^2} \int_0^{\chi_s} d\chi~ \frac{f(\chi) f(\chi_s - \chi) }{a(\chi) f(\chi_s)} ~ \delta\left(\btheta,\chi\right) 
\label{eq:kappa}
\end{align}
\begin{align}
f(\chi) = \left\{
\begin{array}{ccl}
R \sinh(\chi/R) & , & \Omega_C < 0 \\
\chi & , & \Omega_C =0 \\
R \sin(\chi/R) & , & \Omega_C >  0 \\
\end{array}
\right.
\end{align}
where $\chi$ is the coordinate distance, $\chi_s$ is the coordinate distance to the source at redshift $z_s$, $a(\chi)$ is the expansion parameter and $\delta\left(\btheta,\chi\right)$ is the overdensity ($(\rho-\overline{\rho})/\overline{\rho}$).  In the non-flat case, $\Omega_C = 1 - \Omega_m - \Omega_\Lambda$ and $R = \frac{c}{H_o\sqrt{\Omega_C}}$.
 For our purposes here this approximation is more than accurate enough.  

The expected power spectrum and cross-correlation of the deflection potential can be easily calculated using the Fourier space Limber equation in the weak lensing and flat sky 
approximations \citep{Kaiser92}
\begin{align}
C^\kappa_\ell = \frac{9}{4} \frac{H_o^4\Omega_m^2}{c^4} \int\displaylimits_0^{{\rm min}(\chi_s,\chi_s')} d\chi ~ \frac{f(\chi_s-\chi)f(\chi_s'-\chi)}{a(\chi)^2 ~ f(\chi_s) f(\chi_s')} \nonumber \\
\times ~  P_\delta\left( \frac{\ell}{f(\chi)} , z_\chi \right).
\label{eq:Ckappa}
\end{align}
The Poisson equation, $2\kappa = \nabla^2\phi$, relates the potential to the convergence and implies that the power spectrum of $\kappa$  is related to the power spectrum of the potential and displacement by $4 C^\kappa_\ell = \ell^4 C^\phi_\ell  = \ell^2 C^D_\ell $ where $D$ stands for displacement.
In our calculations we use the linear matter power spectrum fit due to \cite{1999ApJ...511....5E} transformed into a nonlinear power spectrum using the method of \cite{peac96}.

Figure~\ref{fig:dC_dz} shows the integrand of Equation (\ref{eq:Ckappa}) as a function of redshift and normalized so that its integral is one, so that one can  see which redshifts contribute most to the signal.  It can be seen that the redshift distribution changes as a function of the source redshift and also the multipole.  We can also see that the self-lensing, lensing by matter at the same redshift as some sources, should be relatively small compared to galaxy lensing.  For sources at $z=3.0$ only 3\% of the signal comes from redshift above $z=2.0$ for $L = 200$.  For this reason we consider it a good approximation to ignore self-lensing here although in the future it might be important to take account of it.

The redshift distribution of the signal does change as a function of source redshift and for this reason it could be possible to measure the evolution of structure formation and angular size distance by breaking the \lya\ data cube into redshift bins and measuring the lensing in each one separately.  This will be discussed further in section~\ref{sec:estimates}.

\section{Noise estimates}
\label{sec:estimates}

\begin{figure}
 \includegraphics[width=\columnwidth]{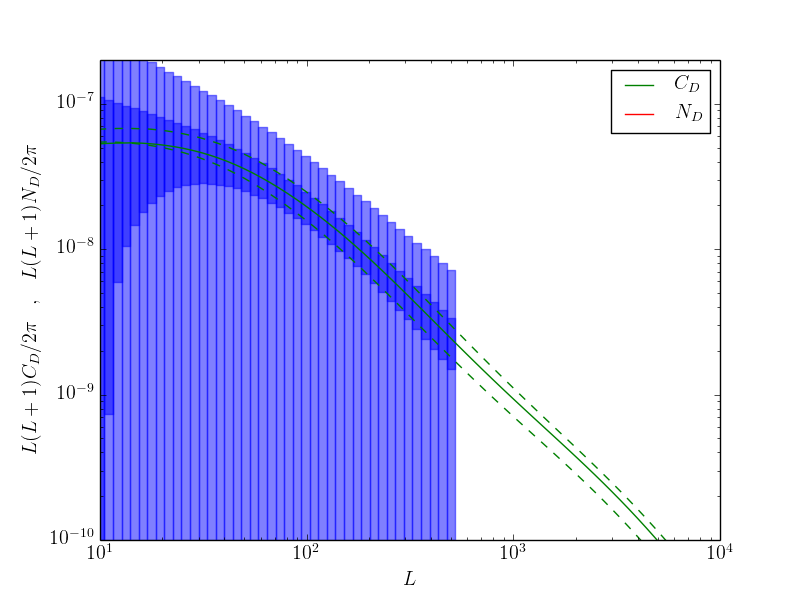}
 \caption{The lensing displacement power spectrum  and its estimated errors for a \lya\ forest survey with the BOSS parameters listed in Table~\ref{table:surveytab}. The smaller errors are for a redshift range $\Delta z =0.5$ and the larger for $\Delta z = 0.1$.  For this survey,  $N_D$ (the noise per mode, red line)  is off the range of the plot.  The green curves labelled $C_D$ show the expected power spectra, with the solid curve being for $z_s=2.5$.  For comparison, the upper dashed green curve shows results for $z_s=3$ and the lower one for $z_s=2$.}
 \label{fig:BOSS}
\end{figure}

\begin{figure}
 \includegraphics[width=\columnwidth]{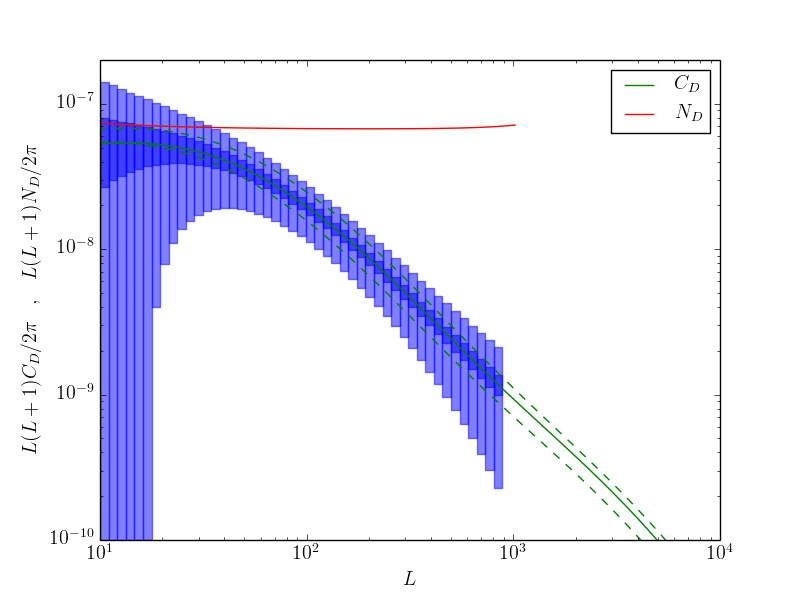}
 \caption{Same as figure~\ref{fig:BOSS} except for the eBOSS parameters listed in table~\ref{table:surveytab}.  The red curve is for $\Delta z =0.5$ while the $\Delta z =0.1$ is off the 
 top of the figure.}
 \label{fig:eBOSS}
\end{figure}

\begin{figure}
 \includegraphics[width=\columnwidth]{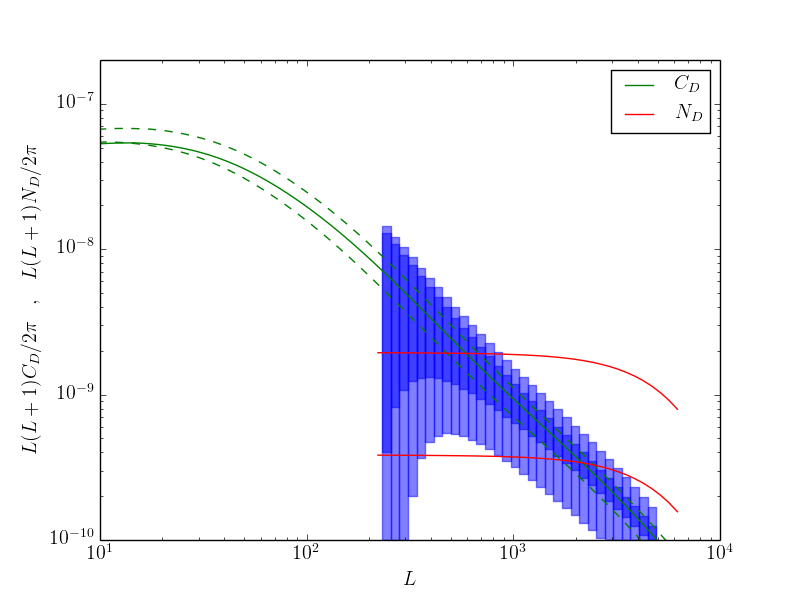}
 \caption{Same as figure~\ref{fig:BOSS} except for the CLAMATO parameters listed in table~\ref{table:surveytab}.  The two red curves are the noise in each mode, $N_D$, for the $\Delta z =0.5$ (lower) and $\Delta z = 0.1$ (higher) cases.  Where $N_D$ is below $C_D$ high fidelity maps of the lensing convergence is possible.}
 \label{fig:CLAMATO}
\end{figure}

\begin{figure}
 \includegraphics[width=\columnwidth]{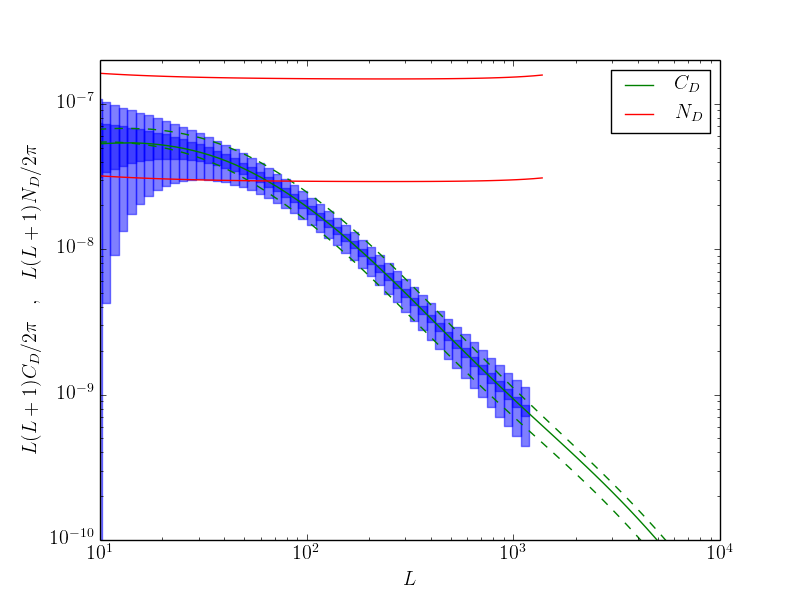}
 \caption{Same as figure~\ref{fig:CLAMATO} except for the WEAVE parameters listed in table~\ref{table:surveytab}.}
 \label{fig:WEAVE}
\end{figure}

\begin{figure}
 \includegraphics[width=\columnwidth]{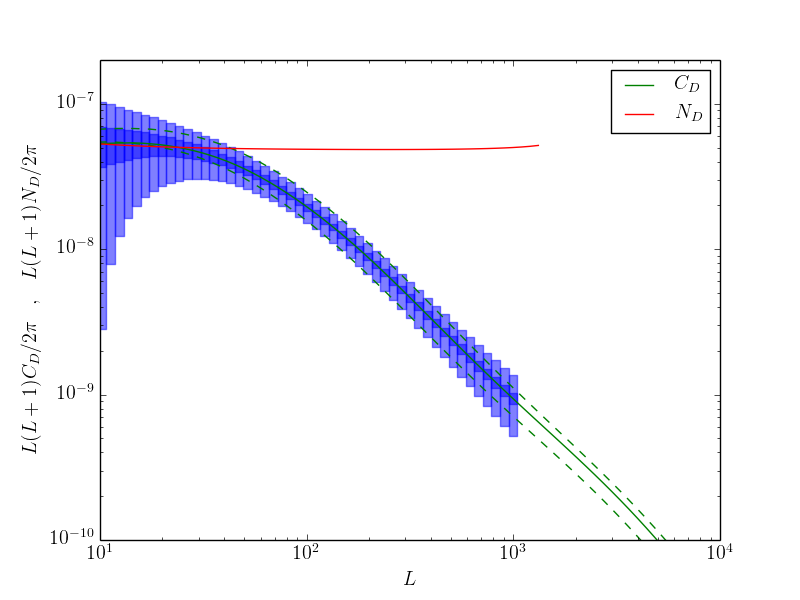}
 \caption{Same as figure~\ref{fig:CLAMATO} except for the DESI parameters listed in table~\ref{table:surveytab}.  The lowest $N_D$ curve is just visible at the top of the plot.}
 \label{fig:DESI}
\end{figure}

\begin{figure}
 \includegraphics[width=\columnwidth]{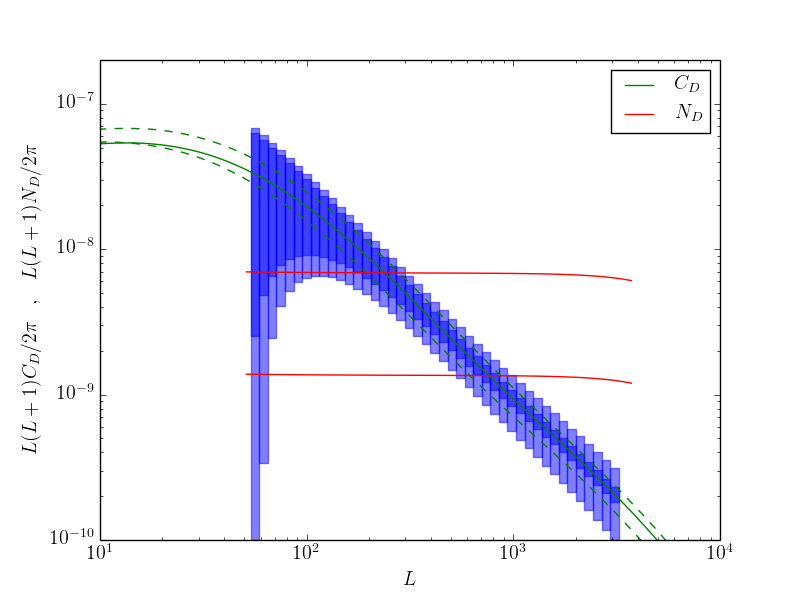}
 \caption{Same as figure~\ref{fig:CLAMATO} except for the Subaru PFS parameters listed in table~\ref{table:surveytab}.}
 \label{fig:SubaruPFS}
\end{figure}

\begin{figure}
 \includegraphics[width=\columnwidth]{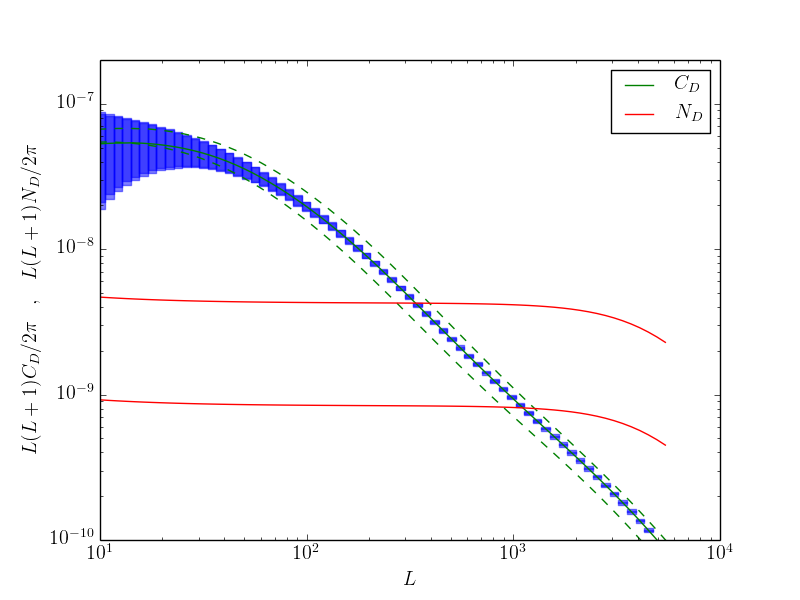}
 \caption{Same as figure~\ref{fig:CLAMATO} except for the MSE parameters listed in table~\ref{table:surveytab}.}
 \label{fig:MSE}
\end{figure}

Using the estimated parameters given in table~\ref{table:surveytab} we have calculated the lensing error, equation~(\ref{eq:Nellj}), and the power spectrum error, equation~(\ref{eq:DeltaC}), for each of the surveys.  In each case we use $\delta z = 0.001$ for the pixel size and a constant signal-to-noise per pixel.  We assume a standard redshift of 2.5 and redshift ranges of $\Delta z =0.5$ and $\Delta z = 0.1$.  The results are displayed in terms of displacement power spectrum for aesthetic reasons.   Plots of the convergence or lensing potential power spectra would be equivalent. We also show the total signal-to-noise, equation~(\ref{eq:ttotal_sig_to_noi}), for each case in table~\ref{table:total_s_to_n}.

Figure~\ref{fig:BOSS} shows the estimated errors for the estimated parameters of the BOSS survey, from Table~\ref{table:surveytab}.  Because of the relatively low density of backlights on the sky and the relatively low signal-to-noise spectra this is the lowest lensing signal-to-noise case we investigate, but even here gravitational lensing could be detectable and some power spectrum information accessible in these data.  If the spectra are broken up into bins of $\Delta z =0.1$ the power spectrum would not be measurable in individual bins.

The situation is more promising for eBOSS, figure~\ref{fig:eBOSS}, where the density of sources is higher.  In this case, if the full range redshift range of $\Delta z = 0.5$ is used a good power spectrum should be recoverable and even a moderately good spectrum within individual $\Delta z = 0.1$ bins although probably not good enough to measure the expected evolution in the power spectrum.  For BOSS and eBOSS, the expected lensing power spectrum never exceeds the noise per mode, $N_D(L)$ and for this reason it would not be possible to create a high fidelity map of the convergence.  For comparison we have included the expected lensing power spectrum at $z=2$ and $z=3$ to give a sense of whether the evolution of the power spectrum could be detected.

The CLAMATO survey will cover a smaller area of sky with a higher density of backlights.   From figure~\ref{fig:CLAMATO} it can be seen that the range in multipole is shifted to smaller scales relative to the BOSSs.  In this case the noise per mode is below the expected power spectrum for a range of $L$.  This indicates that it would be possible to map the convergence with a resolution given be the scale at which $N_D(L)$ and $C_D(L)$ intersect.  Even when the spectra are divided up into bins of $\Delta z = 0.1$ good signal-to-noise is retained indicating that some tomographic information should be recoverable.  The uncertainty in the power spectrum measurement where $N_D(L) < C_D(L)$ is dominated by sample variance owing to the small survey area.

WEAVE and DESI surveys will have larger areas than CLAMATO and higher backlight densities than the BOSSs.  One can see in figures~\ref{fig:WEAVE} and \ref{fig:DESI} that this results in an excellent power spectrum measurement at least within a wide redshift range.  $N_D(L)$ is not low enough in these cases to map the convergence except perhaps for WEAVE at the largest scales.  This is largely a consequence of the lower backlight density relative to CLAMATO.  Subaru PFS will have an area smaller than WEAVE or DESI, but larger that CLAMATO with a backlight density that is similar to CLAMATO.  Here lens mapping will be possible within small redshift ranges and tomographic information should be accessible (figure~\ref{fig:SubaruPFS}).  Finally, figure~\ref{fig:MSE} shows that a very ambitious MSE survey would produce an excellent measurement of the power spectrum in bins as small as $\Delta z =0.1$ and large high fidelity maps with a resolution of $\sim 1\degree$.

The cumulative signal to noise ratio is useful statistic to gauge the overall significance of detection and the precision possible in power
spectrum normalization from observations. We define this quantity in Appendix A, and in Table 2. we show calculated values for the different surveys we have considered. We can see that forest lensing is likely to be highly competitive with galaxy shear surveys.

 \begin{table*}
\begin{tabular}{lccccccc}
\hline
                           & BOSS DR12 &  eBOSS  & CLAMATO   & WEAVE     & DESI      & Subaru PFS & MSE    \\\hline
$\Delta z = 0.1$  & 1.3             & 3.7       & 2.7             &  7.3          & 7.0        &  4.2           &  53 \\
$\Delta z = 0.5$  & 6.0             & 15.9     & 7.7             &   28.0       & 28.8      & 12.5          &  162\\
\hline
\end{tabular}
\caption{ The estimated cumulative signal-to-noise for the lensing power spectrum, $\mathcal{S}$, (defined in the appendix) for different \lya\ surveys and redshift bin width.   For comparison a galaxy shear survey that covers the whole sky, has  intrinsic galaxy ellipticities of 0.3, a source redshift of 0.9, 30 galaxies per arcmin$^2$ and no other noise would have an $\mathcal{S}$ of $\sim 12$. 
\label{table:total_s_to_n}
}
\end{table*}

\section{conclusion}
\label{sec:conclusion}

Weak gravitational lensing should be measurable with high precision using the \lya\ forest data from surveys 
of the scale and quality being done today and in the near future.  Our estimates show that the lensing power spectrum could be measured for source redshifts of $z\simeq 2-3$ in this way and that the evolution of
clustering over this range should be measurable.  This would provide weak lensing information for a range of source redshifts that is not accessible with galaxy shear surveys or lensing of the CMB and allow for probing early dark energy and certain alternative gravity theories.

Cross-correlating \lya\ lensing maps with foreground surveys or galaxy lensing surveys boosts the signal-to-noise and provides access to unique information.
With the ability to map the convergence with high fidelity comes the possibility of another kind of cosmological probe.   The convergence map for one redshift and the convergence map for another will have the same foreground matter, but different $\chi_s$ in equation~(\ref{eq:kappa}) which is a function of cosmological parameters.  With additional redshift information on the foreground, one might be able to constrain the foreground density and the cosmology simultaneously.  Such a measurement would not suffer from cosmic variance and be a more direct measure of the angular size / redshift relation.  This possibility will be further investigated in the future.

\section*{Acknowledgments}
We thank Andreu Font, Mat Pieri and K.G. Lee for their help with understanding
the various samples of observational data and for their willingness to
give approximate estimates of future data properties.
AR and RBM have been supported partly through project GLENCO, funded under the European Seventh Framework Programme, Ideas, Grant Agreement n. 259349. RACC is supported by NASA ATP award NNX17AK56G.

\bibliographystyle{mn2e}
\bibliography{references}
\onecolumn
\appendix

\section{estimated error in power spectrum normalization}

To forecast how well the power spectrum can be recovered we start with the likelihood for Fourier modes with a Gaussian prior
\begin{align}
\mathcal{L} = \frac{1}{(2\pi)^N \prod_\ell \sqrt{N_\ell} \prod_\ell \sqrt{C_\ell} } \exp\left[ - \frac{1}{2} \sum_\ell \frac{(\hat{a}_\ell - a_\ell)^2}{N_\ell}  \right] \exp\left[ - \frac{1}{2} \sum_\ell \frac{ a_\ell^2}{C_\ell}  \right].
\end{align}
Here $C_\ell$ is the signal power spectrum and $N_\ell$ is the noise per mode.  The actual amplitudes are $a_\ell$ and $\hat{a}_\ell$.  In the real case with an irregular survey shape, non-isotropic noise and randomly distributed backlights there will be cross-correlations between different modes and this must be done more carefully, but for our purposes we will consider then uncorrelated.

We can calculate, marginalizing this over the mode values by integrating over the $a_\ell$'s
\begin{align}
\hat{\mathcal{L}} = \frac{1}{(2\pi)^{N/2} \prod_\ell \sqrt{ N_\ell + C_\ell }}  \exp\left[ - \frac{1}{2} \sum_\ell \frac{ \hat{a}_\ell^2}{C_\ell + N_\ell}  \right]
\end{align}
The error for a parameter is often estimated with the Fisher information for that parameter which is the average of the second derivative of the log of the likelihood with respect to that parameter.   In this case we will take the normalization of the power spectrum $A$ to be our parameter.  The information for this parameter is  
\begin{align}
\label{eq:ttotal_sig_to_noi}
\mathcal{S}^2 \equiv - \left\langle \frac{\partial^2 \ln \hat{\mathcal{L}} }{\partial^2 A} \right\rangle 
= \frac{1}{2} \sum_\ell \frac{C_\ell^2}{\left( N_\ell + C_\ell\right)^2} 
\simeq \frac{A_{\rm survey}}{2(2\pi)^2} \int d^2\ell ~\frac{C_\ell^2}{\left( N_\ell + C_\ell\right)^2} 
= \frac{A_{survey}}{4\pi} \int d\ell \ell ~\frac{C_\ell^2}{\left( N_\ell + C_\ell\right)^2} 
\end{align}
In converting from the sum to the integral the density of modes in a finite size survey is approximated as $\Delta\ell \simeq 2\pi / \sqrt{A_{\rm survey}}$.
This can be applied to the whole spectrum, in which case this is the square of signal-to-noise for the power spectrum normalization.  It can also be applied to bands in $\ell$ which, if the noise and spectra are approximated as linear within the bands, is equation~\ref{eq:DeltaC} for the lensing spectrum .

\end{document}